\documentclass[showpacs,preprintnumbers,amsmath,amssymb]{revtex4}
\usepackage{dcolumn}
\usepackage{bm}

\begin{document}

\title{Quantum Black Holes. Black Hole Temperature without a Black Hole.}

\author{Victor Berezin}
\affiliation{Institute for Nuclear Research \\
        Russian Academy of Sciences\\
        Moscow, 117312, 60-October Anniversary pr.,7-a}
\email{berezin@ms2.inr.ac.ru}

\begin{abstract}
The model is constructed, some features of which comes from quantum thin
dust shells and is, in fact, an extension of the "no hair" property of classical black hole
on a quantum level. It appears that the proposed classical analog of quantum black hole
is heated, the temperature being exactly the Hawking's temperature.
\end{abstract}

\maketitle

\section{Quantum shells.}

In series of papers \cite{1} a quantum mechanics for spherically
symmetric thin dust shells has been developed. Here we write out
only some necessary results. It was shown that the Wheeler-DeWitt
equation is reduced in this case to the following stationary
Schroedinger equation in finite differences:
\begin{equation}
\label{Psi}
\Psi(m, m_{in}, S + i\zeta) + \Psi(m, m_{in}, S - i\zeta) = \frac{F_{in} + F_{out} - \frac{M^2}{4 m^2 S}}{\sqrt{F_{in}} \sqrt{F_{out}}}
\Psi (m, m_{in}, S),
\end{equation}
where $m = m_{out} = m_{tot}$ - the total mass of the system,
$m_{in}$ - the Schwarzschild mass inside, $M$ is the bare mass of
the shell, $S = \frac{R^2}{4 G^2 m^2}$ ($R$ - radius, $G$ -
gravitational constant), $F = 1 - \frac{2 G m}{R},\, \zeta =
\frac{m_{Pl}^2}{2 m^2}$ ($m_{Pl} = \sqrt{\frac{\hbar c}{G}}$ is the
Planckian mass and we use units with $\hbar = c = k = 1$, $\hbar$ -
Planck constant, $c$ - speed of light, $k$ - Bolzmann constant). By
investigation of wave functions in the vicinity of singular points
(infinities and singularities) and around the branching points
(apparent horizons) the following discrete mass spectrum for bound
states was found $(\Delta m = m_{out} - m_{in})$:
\begin{eqnarray}
\label{spectrum}
\frac{2 (\Delta m)^2 - M^2}{\sqrt{M^2 - (\Delta m)^2}} &=&
\frac{2 m_{Pl}^2}{\Delta m + m_{in}}\, n \,, \nonumber \\
M^2 - (\Delta m)^2 &=& 2 (1 + 2 p)\, m_{Pl}^2 \,,
\end{eqnarray}
where $n$ and $p \ge 0$ are integers. The appearance of two quantum
numbers instead of one in conventional quantum mechanics is due to
the nontrivial causal structure of the complete Schwarzschild
manifold.

The above spectrum is not universal in the sense that the
corresponding wave functions form a two-parameter family
$\Psi_{n,p}$. But for the quantum Schwarzschild black hole we expect
a one-parameter family of solutions, because quantum black holes
should not have "no hairs", otherwise there will be no smooth
classical limit. This means that our spectrum is not a quantum black
hole spectrum, and corresponding quantum shells do not collapse
(like an electron in hydrogen atom). Physically, it is quite
understandable, because the radiation was not included into
consideration. The energy of radiation is also quantized, but, and
this is crucial, the energies of quanta are not equal to the level
splitting in the shell discrete spectrum, Eqn.(\ref{spectrum}). As a
result, the quantum gravitational collapse proceeds via production
new shells, thus increasing the inner mass $m_{in}$ inside the
primary shell. Such a process can go in many different ways, so, it
is the quantum collapse that appears to be the origin of the black
hole nonzero entropy. But how could quantum collapse be stopped? The
natural limit is the transition from a black hole-like shell to a
wormhole-like shell by crossing an Einstein-Rosen bridge, since such
a transition requires (at least in a quasi-classical regime)
insertion of infinitely large volume, which probability is, of
course, zero. Computer simulations show that the process of quantum
gravitational process stops when the principal quantum number
becomes zero, $n = 0$.

The point $n = 0$ in our spectrum is very special. In this state the
shell does not "feel" not only the outer region (what is natural for
the spherically symmetric configuration), but it does not know
anything about what is going on inside. It "feels" only itself. Such
a situation reminds the "no hair" property of a classical black
hole. Finally, when all the shells (both the primary one and newly
produced) are in the corresponding states $n_i = 0$, the whole
system does not "remember" its own history. And it is this "no
memory" state that can be called "the quantum black hole". Note,
that the total masses of all the shells obey the relation
\begin{equation}
\label{miMi}
\Delta m_i = \frac{1}{\sqrt{2}} M_i.
\end{equation}
The subsequent quantum Hawking's evaporation can proceed via some
collective excitations and formation, e.g., of a long chain of
microscopic semi-closed worlds.

\section{Classical analog of quantum black hole.}

The final state of quantum gravitational collapse, the quantum black
hole, can be viewed as some stationary matter distribution.
Therefore, we may hope that for massive enough quantum black hole
such a distribution is described approximately by a classical static
spherically symmetric perfect fluid with energy density
$\varepsilon$ and pressure $p$ obeying classical Einstein equations.
This is what we call a classical analog of a quantum black hole. Of
course, in such a case the corresponding classical distribution has
to be very specific. To study its main features let us consider the
situation in more details.

Any static spherically symmetric metric can be written in the form
\begin{equation}
\label{ds}
d s^2 = e^{\nu} d t^2 - e^{\lambda} d r^2 - r^2 (d \theta ^2 +
\sin^2{\theta} d \varphi ^2).
\end{equation}
Here $r$ is the radius of a sphere with the area $S = 4 \pi r^2, \nu
= \nu (r), \lambda = \lambda (r)$. The Einstein equations are (prime
denotes differentiation in $r$):
\begin{eqnarray}
\label{Ee}
- e^{- \lambda} \left(\frac{1}{r^2} - \frac{\lambda^{\prime}}{r} \right)
+ \frac{1}{r^2} &=& 8 \pi G \varepsilon \,, \nonumber \\
- e^{- \lambda} \left(\frac{1}{r^2} + \frac{\nu ^{\prime}}{r} \right) +
\frac{1}{r^2} &=& - 8 \pi G p \,, \nonumber \\
- \frac{1}{2} \left( \nu^{\prime \prime} + \frac{\nu ^{\prime^2}}{2} +
\frac{\nu ^{\prime} - \lambda ^{\prime}}{r} -
\frac{\nu ^{\prime} \lambda ^{\prime}}{2} \right) &=& - 8 \pi G p \,.
\end{eqnarray}
We see that there are three equations for two unknown functions of
one variable, namely, $\nu(r), \lambda(r), \varepsilon(r)$ and
$p(r)$. But, even we would know an equation of state for our perfect
fluid, $p = p(\varepsilon)$, the closed (formally) system of
equations would have too many solutions. We need, therefore, some
selection rules in order to single out the classical analog of
quantum black hole. Surely, the "no hair" feature should be the main
criterium. Thus, we have to adjust our previous definition of the
"no memory" state to the case of a continuum matter distribution.
For this, let us integrate the first of Eqns.(\ref{Ee}):
\begin{equation}
\label{F}
e^{- \lambda} = 1 - \frac{2 G m(r)}{r},
\end{equation}
where
\begin{equation}
\label{mr}
m(r) = 4 \pi \int\limits_0^r \varepsilon \tilde r^2 d \tilde r
\end{equation}
is the mass function that must be identified with $m_{in}$. Now, the
"no memory" principle is readily formulated as the requirement, that
$m(r) = a r^2$  \cite{2}, i.e.,
\begin{equation}
\label{constlambda} e ^{- \lambda} = 1 - 2 G a = const .
\end{equation}
Note, that in static case, the inverse metric coefficient $e^{-
\lambda}$ is an invariant which in the general spherically symmetric
space-time reads as $\Delta = - e^{- \lambda} = g^{ik} R_{, i} R_{,
k}$ and is nothing more but a squared normal vector to the surface
of constant radius $R(x^i) = R(t, q) = const$. We can also introduce
a bare mass function $M(r)$ (the mass of the system inside a sphere
of radius $r$ without the gravitational mass defect).

\begin{equation}
\label{Mr}
M(r) = 4 \pi \int\limits_0^r \varepsilon (\tilde r)
 e^{\frac{\lambda}{2}}(\tilde r) \tilde r^2 d \tilde r =
 \frac{a r}{\sqrt{1 - 2 G a}} \,.
\end{equation}
The remaining two equations (\ref{Ee}) can now be solved for $p(r)$
and $e^{\nu}(r)$. The general solution is rather complex, but the
correct non-relativistic limit for the pressure $p(r)$ (we are to
reproduce the famous equation for hydrostatic equilibrium) has only
the following one-parameter family:
\begin{equation}
\label{pres}
p(r) = \frac{b}{4 \pi r^2}\, ,
\end{equation}
where
\begin{equation}
\label{b}
b = \frac{1}{G} \left( 1 - 3 G a - \sqrt{1 - 2 G a} \sqrt{1 - 4 G a} \right) \,.
\end{equation}
We see that the solution exists only for $a \le \frac{1}{4 G}$, then
$b \le a$. The physical meaning of these inequalities is that the
speed of sound cannot exceed the speed of light, $v_{sound}^2 =
\frac{b}{a} \le 1 = c^2$, the equality being reached just for $a = b
= \frac{1}{4 G}$. Finally, for the temporal metric coefficient
$g_{00} = e^{\nu}$ we get:
\begin{equation}
\label{nu}
e^{\nu} = C_0 r^{\frac{4 b}{a + b}} = C_0 r^{2 G \frac{a + b}{1 - 2 G a}} \,.
\end{equation}
Thus, demanding the "no memory" feature and existence of the correct
non-relativistic limit, we obtained the two-parameter family of
static solutions. But we need a one-parameter family, so we have to
continue our search. Let us investigate the obtained space-time
manifolds more thoroughly, especially in the vicinity of the
apparently singular point $r = 0$, and calculate the corresponding
curvature (Riemann's) tensor $R^{\mu}_{\nu \lambda \sigma}$. The
nonzero components are
\begin{eqnarray}
\label{Riemann}
R^0_{101} &=& 2 \frac{b (a - b)}{(a + b)^2} \frac{1}{r^2} \,; \;\;
R^0_{202} = 2 \frac{b}{a + b} (1 - 2 G a)\, ; \;\;
R^0_{303} = R^0_{202} \sin^2 {\theta}\,; \nonumber \\
R^1_{010} &=& 2 C_0^2 \frac{b (b - a)}{(a + b)^2}
(1 - 2 G a) r^{\frac{2 (b -a)}{a + b}} \,; \nonumber \\
R^2_{020} &=& 2 C_0^2\frac{b}{a + b} (1 - 2 G a)
r^{\frac{2 (b -a)}{a + b}} \,; \;\; R^2_{323} = 2 G a \sin^2 {\theta} \,;
\nonumber \\
R^3_{030} &=& 2 C_0^2 \frac{b}{a + b} (1 - 2 G a) r^{\frac{2 (b - a)}{a + b}} \,;\;\;
R^3_{232} = 2 G a \,.
\end{eqnarray}
Evidently, for $b < a$ the Riemann tensor (\ref{Riemann}) is
divergent at $r = 0$, so, the corresponding space-times have the
real singularity. But, if $a = b = \frac{1}{4 G}$ we are witnessing
a miracle, the (before) divergent components become zero, and the
remaining nonzero ones equal
\begin{eqnarray}
\label{Riemann1}
R^0_{202} &=& - (1 - 2 G a) = - \frac{1}{2}\,, \;\;  \left(R^2_{020} =
\frac{1}{2} C_0^2 \right)\, ; \nonumber \\
R^0_{303} &=& - (1 - 2 G a) = - \frac{1}{2}\,, \;\;  \left( R^0_{030} =
\frac{1}{2} C_0^2 \right)\, ; \nonumber \\
R^2_{323} &=& 2 G a \sin^2 {\theta} = \frac{1}{2} \sin^2{\theta}\,,\;\;
\left( R^3_{232} = \frac{1}{2} \right) \,,
\end{eqnarray}
and the only nonzero component of the Ricci tensor $R_{\mu \nu} ( =
R^{\alpha}_{\mu \alpha \nu})$ equals to
\begin{equation}
\label{Ricci}
R_{00} = C_0^2 .
\end{equation}
Thus, demanding, in addition to the previous two very natural
requirements, the third one (also natural), namely, the absence of
the real singularity at $r =0$, we arrive at the following
one-parameter family to the Einstein equations (\ref{Ee}):
\begin{eqnarray}
\label{soln}
g_{00} &=& e^{\nu} = C_0^2 r^2 , \nonumber \\
g_{11} &=& - e^{\lambda} = - \sqrt{2} , \nonumber \\
\varepsilon &=& p = \frac{1}{16 \pi G r^2} .
\end{eqnarray}
So, the equation of state of our perfect fluid is the stiffest
possible one. The constant of integration $C_0$ can be determined by
matching the interior and exterior metrics at some boundary radius
$r = r_0$. Let us suppose that for $r > r_0$ the space-time is
empty, so, the interior should be matched to the Schwarzschild
metric, labeled by the mass parameter $m$. Of course, to compensate
the jump in the pressure $\Delta p \,( = p(r_0) = p_0)$ we must
include in our model some surface tension $\Sigma$. It is easy to
check, that
\begin{eqnarray}
\label{match}
C_0^2 &=& \frac{1}{2 r_0^2}\, ; \;\; \Delta p = \frac{2 \Sigma}{\sqrt{2} r_0} \,;
\nonumber \\
e^{\nu} &=& \frac{1}{2} \left(\frac{r}{r_0} \right)^2 ;\;\; p_0 = \varepsilon _0
= \frac {1}{16 \pi G r_0^2} \,;\nonumber \\
m &=& m_0 = \frac{r_0}{4 G}\, .
\end{eqnarray}
Note, that the bare mass $M = \sqrt{2} m$, the relation is exactly
the same as for the shell "no memory" state (\ref{miMi}), and $r_0 =
4 G m_0$, so, the size of our analog of quantum black hole is twice
as that of classical black hole. But how about the special point in
our solution, $r = 0$? It is not a trivial coordinate singularity,
like in a three-dimensional spherically symmetric case, because
\begin{equation}
\label{ds0}
ds^2 (r = 0) = 0.
\end{equation}
This looks rather like an event horizon. To clear the point we
consider the radial geodesic motion in the space-time with the
metric
\begin{equation}
\label{ourds}
ds^2 = \frac{1}{2} \left(\frac{r}{r_0} \right) d t^2 - 2 d r^2 -
r^2 (d \theta ^2 + \sin^2 {\theta} d \varphi ^2 ) .
\end{equation}
The calculations are very simple, the falling bounded geodesics are
described by the following function $r(t)$, or $r(\tau)$ for the
proper time parameter $\tau$ :
\begin{eqnarray}
\label{geod}
r(\tau) &=& \frac{1}{\sqrt{2}} \sqrt{2 r_1^2 - \tau ^2} ; \nonumber \\
\tau &=& \sqrt{2}r_1 \tanh {\frac{t}{2 r_0}} ; \nonumber \\
r(t) &=& \frac{r_1}{\cosh {\frac{t}{2r_0}}} ,
\end{eqnarray}
where we put $r = r_1 < r_0$ for $\tau = t =0$
($\frac{dr}{d\tau}(r_1) = \frac{dr}{dt}(r_1) = 0$). We see that the
surface $r =0$, indeed, behaves like an event horizon (and, at the
same time, a Killing horizon). Investigation of non-radial geodesics
shows an infinite spiralling when approaching zero radius surface,
thus confirming its horizon nature. Moreover, the two-dimensional
part of the metric (\ref{ourds}), i.e., $(t-r)$-surface, is locally
flat, what can easily be proven by making the following coordinate
transformation $(t,r) \to (\eta, x)$:
\begin{equation}
\label{lf}
\eta = \sqrt{2} r \sinh {\frac{t}{2 r_0}}\, , \;\;
x = \sqrt{2} r \cosh {\frac{t}{2 r)}} \,.
\end{equation}
This resembles the Rindler's transformation in two-dimensional flat
Minkowski space-time.

\section{Rindler space-time.}

The Rindler space-time is obtained by transforming the
two-dimensional Minkowski space-time from the ordinary coordinates
$(\eta,x)$ and metric $ds^2 = d\eta ^2 - dx^2$ related to the set of
inertial observers, to the so-called Rindler coordinates and metric
\begin{eqnarray}
\label{Rindler}
\eta &=& \frac{1}{a} e^{a \xi} \sinh {a \eta} \,,\;\;
x = \pm \frac{1}{a} e^{a \xi} \cosh {a \eta} \;\; (x \ge 0), \nonumber \\
&& -\infty < t < \infty , \;\;  - \infty < \xi < \infty , \nonumber \\
&& d s^2 = e^{2 a \xi} (d t^2 - d\xi ^2) .
\end{eqnarray}
Thus, the Rindler space-time is static and locally flat but differs
from the two-dimensional Minkowski space-time globally, because it
covers only one half of the latter and, in addition, possesses the
event horizons at $\eta = \pm x \;(t = \pm \infty, \,\xi = const)$.
The Rindler observers $\xi = const$ undergo nonzero constant
acceleration. The norm of the acceleration vector $a^{\mu}$ equals
\begin{equation}
\label{acc}
\alpha = \sqrt{|a^{\mu} a_{\mu}|} = a e^{- a \xi}.
\end{equation}
Existence of the horizons has a very important consequence.
W.G.Unruh showed \cite{3} that the quantum theory of scalar field in
the Rindler space-time is, in fact, the finite temperature quantum
field theory, and the value of the Unruh's temperature is
\begin{equation}
\label{tunruh}
T_U = \frac{a}{2 \pi} .
\end{equation}
We see, that this temperature is proportional to the acceleration of
the Rindler observer sitting at $\xi =0$ with $g_{00} = 1$. But, all
these observers are equivalent (we can always shift the spatial
coordinate $\xi \to \xi - \xi_0$). The temperature value is not an
invariant but it is a temporal component of a heat vector. This
means that each observer measures the Unruh temperature when using
its proper time $\tau \;(ds = d\tau)$. If the same observer uses the
local clocks that show the local time $t \;(ds = \sqrt{g_{00}} dt)$,
the local temperature measured by him equals
\begin{equation}
\label{tloc}
T_{loc} = \frac{T_U}{\sqrt{g_{00}}} = \frac{a}{2 \pi} e^{- a \xi}
= \frac{\alpha}{2 \pi} .
\end{equation}
We know from the university course of thermodynamics (se, e.g.,
\cite{3}) that the condition for thermal equilibrium in static
space-times is $T_{loc} \sqrt{g_{00}} = const$. We can introduce an
"apparent temperature" $T_{app}$ which is the local temperature of
an observer sitting at $\xi = \xi_2$, "seen" by an observer at $\xi
= \xi_1$:
\begin{equation}
\label{tapp}
T_{app} (\xi_2, \xi_1) = \frac{\sqrt{g_{00}(2)}}{\sqrt{g_{00}(1)}} T_{loc}(\xi_2) .
\end{equation}
Then the condition for thermal equilibrium can be formulated as
$T_{loc}(\xi_1) = T_{app}(\xi_2)$ for all values  $\xi = \xi_2$.
Thus, the Rindler observers are in thermal equilibrium with each
other. And the question arises: is the Rindler space-time unique in
this sense? To answer it, consider some general two-dimensional
static space-time with a metric
\begin{equation}
\label{uni}
d s^2 = e^{\nu} d t^2 - d \rho ^2 = e^{\nu} d t^2 - e^{\lambda} d q ^2 .
\end{equation}
Note, that in the Rindler case $\rho = \frac{1}{a} e^{a \xi},
e^{\nu} = \frac{\rho ^2}{a^2} = g_{00}$. The static observer in the
metric (\ref{uni}) undergoes a constant acceleration with the
invariant $\alpha = \frac{1}{2} |\frac{d \nu}{d \rho}| = \frac{1}{2}
|\frac{d \nu}{d q}| e^{- \frac{\lambda}{2}}$, and the (local)
Rindler parameter $a (\rho)$, which is now called "the surface
gravity $\kappa$", equals
\begin{equation}
\label{kappa}
\kappa = \frac{1}{2} \left|\frac{d \nu}{d q} \right| e^{\frac{\nu - \lambda}{2}} =
\frac{1}{2} \left|\frac{d \nu}{d \rho} \right| e^{\frac{\nu}{2}} \, .
\end{equation}
The thermal equilibrium condition requires $\kappa = const$,
therefore, $g_{00} = C \rho ^2$, and this proves that the Rindler
space-time is the only one which static observers are in the mutual
thermal equilibrium.

 \section{Topological temperature.}

The fact that an accelerated observer "sees" particles while moving
in the empty (vacuum) Minkowski space-time, was known long ago to
quantum field theorists. The physical reason for this phenomenon is
obvious: particle creation is caused by the same forces that cause
the particle detector's acceleration. From the quantum field theory
point of view, the vacuum of an accelerated observer (=detector) is
different from that of an inertial observer. And the empty space of
the latter appeared filled with particles to the first one. Due to
Unruh's discovery we know that the observer moving with constant
acceleration detects particles with Planckian spectrum at the
temperature proportional to the value of this acceleration. But, let
us consider the following Gedanken experiment. Two observers sitting
in the rockets and bringing particle detectors with them, are moving
inertially in the two-dimensional Minkowski space-time. At some
definite moment they switch on the engines and start to move with
equal constant accelerations. Surely, their detectors start to
register particles. Because an amount of fuel in the rockets is
finite, our observers will eventually become inertial again, and let
the durations of their accelerated motion are different. Then, the
number of detected particles will also be different and finite.
Suppose, our observers are well educated and know about the Unruh
effect, so, they are able to compare the Planckian spectrum with
that obtained by them. Evidently, they will find the deviations from
the thermal spectrum, and the longer their accelerations, the
smaller will be these deviations. Thus, the appearance of the
temperature in the Rindler space-time is a global effect - the
acceleration should last infinitely both in the past and in the
future.

Thermodynamically this can be understood as follows. The existence
of the event horizons in the Rindler space-time prevents receiving
any information from the remaining part of a manifold. And the
observers can explain this loss of information by ascribing a
nonzero entropy to the unseen part of a geodesically complete
space-time. And the very appearance of the entropy and the static
character of the Rindler space-time lead us to the notion of thermal
equilibrium and, thus, to the notion of temperature. Therefore, we
see that it is the event horizon that causes the spectrum of
particles detected by the constantly accelerated observer to be the
Planckian one. And the global character of the notion of event
horizon is reflected in the global character of the Unruh's
temperature.

It is amazing, but the Rindler temperature can be calculated without
a thorough investigation of quantum field theory. For this we should
make a Whick rotation to the imaginary Rindler time $\tau = it$,
then the Euclidean Rindler metric
\begin{equation}
\label{ER}
d l^2 = a^2 \rho ^2 d \tau ^2 + d \rho ^2
\end{equation}
can be interpreted as that of a locally flat two-dimensional surface
in polar coordinates (the polar angle is proportional to $\tau$)
provided the imaginary time is periodic, the latter requirement is
quite natural because in terms of Minkowskian time $\eta$ such a
transition reads as $\eta = \frac{\rho}{a} \sinh {(i a t)} = i
\frac{\rho}{a} \sin {\tau}$. The metric (\ref{ER}) describes, in
general, the geometry of a cone embedded into the three-dimensional
flat space with a conical singularity at $\rho = 0$. But, if the
period is $\tau = \frac{2 \pi}{a}$, the singularity disappears and
we obtain the whole plane. In the finite temperature field theories
the temperature is introduced as the inverse period of the imaginary
time, and we see that in our case it is exactly the Unruh
temperature (\ref{tunruh})! We can call the temperature found in
this way "the topological temperature", because the geometry in the
vicinity of a single (but singular) point, determines the properties
of the whole manifold.

\section{Black hole temperature with and without black holes.}

By the Einstein equivalence principle we can extend all we learned
studying Rindler space-times, to the static gravitational fields,
especially to the static spherically symmetric manifolds, because
after fixing spherical angles $\theta$ and $\varphi$ they become, in
fact, the two-dimensional surfaces. Of course, in general these
surfaces are curved, the equivalence principle holds only locally
and, therefore, static observers sitting at different values of
radius will "feel" not only different temperatures, but by no means
they will be in thermal equilibrium with each other. Such a
temperature is observer dependent and cannot be considered as an
intrinsic property of a given space-time. But, we saw that the
Rindler space-time possesses the event horizons what is crucial for
ascribing and entropy and temperature to the manifold itself (or to
its part).

So, we are looking for some examples of spherically symmetric static
manifolds which possess the event horizons. And, of course, these
are the well-known Schwarzschild and Reissner-Nordstrom space-times
outside the corresponding black holes. In both cases the metric can
be written in the form
\begin{eqnarray}
\label{SRN}
d s^2 &=& F d t^2 - \frac{1}{F} d r^2 - r^2 (d \vartheta ^2 +
\sin^2 {\vartheta} d \varphi ^2) , \nonumber \\
&& F = 1 - \frac{2 G m}{r} + \frac{G e^2}{r^2} =
\left(1 - \frac{r_+}{r} \right) \left(1 - \frac{r_-}{r} \right) ,
\end{eqnarray}
where $m$ is the mass, $e$ is the electric charge of
Reissner-Nordstrom black hole, $r_{\pm} = G m \pm \sqrt{G^2 m^2 - G
e^2}$ are, respectively, positions of the event and Cauchy horizons,
and for $e = 0$ the above relations become the parameters of a
Schwarzschild black hole. After the Whick rotation to the imaginary
time $\tau$ the Euclidean two-dimensional surface in the vicinity of
the event horizon $r_+$ is described by the metric
\begin{eqnarray}
\label{ERN}
d l^2 &=& \frac{(r_+ - r_-)^2}{4 r_+^2} \rho ^2 d t^2 + d \rho ^2 , \nonumber \\
&& \rho = \frac{2 r_+}{\sqrt{r_+ - r_-}} \sqrt{r_+ r_-} \ll r_+ .
\end{eqnarray}
Thus, the topological temperature of the Reissner-Nordstrom black
hole is
\begin{equation}
\label{tRN}
T_{top} = \frac{r_+ - r_-}{4 \pi r_+^2} .
\end{equation}
This value equals $\frac{\kappa _{BH}}{2 \pi}$, where $\kappa _{BH}$
is the surface gravity calculated at the black hole event horizon
that enters the first law of thermodynamics for black holes \cite{5}
\cite{6}. And the topological temperature $T_{top}$ in the case of
Schwarzschild black hole is exactly the famous Hawking temperature
\begin{equation}
\label{Ht}
T_H = \frac{1}{8 \pi G m} ,
\end{equation}
obtained by explicit construction of the massless scalar quantum
field theory on the Schwarzschild curved background with specific
boundary conditions at the event horizon \cite{7}. The temperature
$T_H$ is the temperature of the Schwarzschild black hole seen by the
distant static observer (at spatial infinity) for whom $g_{00} = 1$.
Other static observers sitting at the radius $r$, see the black hole
apparent temperature $T_{app} = \frac{T_H}{\sqrt{g_{00}(r)}}$. This
apparent temperature does not coincide with their local Rindler
temperature $T_{loc} = \frac{\kappa (r)}{2 \pi \sqrt{g_{00}}}$ what
indicates that the static black hole is not in thermal equilibrium
with the surrounding vacuum space-time (i.e., there is no heat bath)
and, in fact, evaporates.

At last, let us turn to our model, the classical analog of quantum
black hole. The metric (\ref{ourds}) for the internal part of the
model ($r \le r_0,\, \theta, \phi = const$)is already in the Rindler
form, if we put $\rho = \sqrt{2} r$. So, all static observers are in
thermal equilibrium, and calculation of the topological temperature
is straightforward. But there is one subtle and important point, in
our case the horizon is at zero radius, $r = 0$, which is at the
same time the coordinate singularity of the whole four-dimensional
space-time, therefore, we are not allowed to confine ourselves to
the two-dimensional $(t -r)$-section. And after the Whick rotation
we have to consider the four-dimensional Euclidean space-time where,
in addition to the periodic time $\tau$, there are two spherical
angles, $\theta$ and $\phi$. The regularity condition requires the
period $2 \pi$ for only one of the angles, all other should have the
period $\pi$. Since the period $2 \pi$ is already reserved for the
azimuthal angle $\phi$, the topological temperature in our model
equals
\begin{equation}
\label{tour}
T_{top} = \frac{1}{2 \pi r_0} .
\end{equation}
Remembering now that $r_0 = 4 G m_0$, we get for the classical
analog of quantum black hole
\begin{equation}
\label{tf}
T_{BH} = \frac{1}{8 \pi G m_0}
\end{equation}
and this is exactly the Hawking temperature!

\section{Acknowledgements.}

The author is indebted to the Russian Foundation for Fundamental
Researches for financial support (Grant N 06-02-16-342 a).

\end{document}